\documentclass[11pt]{article}

\usepackage[T1]{fontenc}
\usepackage[utf8]{inputenc} 
\usepackage[english]{babel}

\usepackage{lmodern}
\usepackage{microtype}
\usepackage{graphicx}
\usepackage{booktabs}
\usepackage{array}
\usepackage{tabularx}
\usepackage{ragged2e}
\usepackage{longtable} 
\usepackage{url}
\usepackage[hidelinks]{hyperref}
\usepackage{caption}
\usepackage{subcaption}

\usepackage[margin=1in]{geometry}

\title{Never say never: Exploring the effects of available knowledge on agent persuasiveness in controlled physiotherapy motivation dialogues}

\author{
Stephan Vonschallen$^{1,2,3}$ \and
Rahel H\"ausler$^{2}$ \and
Theresa Schmiedel$^{1}$ \and
Friederike Eyssel$^{3}$
}

\date{} 

\begin{document}
\maketitle
\noindent
$^1$Institute of Business Information Technology, Zurich University of Applied Sciences, Switzerland\\
$^2$Institute for Information Systems, University of Applied Sciences and Arts Northwestern Switzerland\\
$^3$Center for Cognitive Interaction Technology, Bielefeld University, Germany\\[4pt]
\textbf{Corresponding author:} Stephan Vonschallen (\texttt{stephan.vonschallen@zhaw.ch})

\begin{abstract}

Generative Social Agents (GSAs) are increasingly impacting human users
through persuasive means. On the one hand, they might motivate users to
pursue personal goals, such as healthier lifestyles. On the
other hand, they are associated with risks like manipulation
and deception, which are induced by limited control over probabilistic
agent outputs. However, as GSAs manifest communicative patterns based on
available knowledge, their behavior may be regulated through their
access to such knowledge. Following this approach, we explored
persuasive ChatGPT-generated messages in the context of human-robot
physiotherapy motivation. We did so by comparing ChatGPT-generated
responses to predefined inputs from a hypothetical physiotherapy
patient. In Study 1, we qualitatively analyzed 13 ChatGPT-generated
dialogue scripts with varying knowledge configurations regarding
persuasive message characteristics. In Study 2, third-party observers
(\emph{N} = 27) rated a selection of these dialogues in terms of the
agent's expressiveness, assertiveness, and persuasiveness. Our findings
indicate that LLM-based GSAs can adapt assertive and expressive personality
traits-- significantly enhancing perceived persuasiveness. Moreover,
persuasiveness significantly benefited from the availability of
information about the patients' age and past profession, mediated by
perceived assertiveness and expressiveness. Contextual knowledge about
physiotherapy benefits did not significantly impact persuasiveness,
possibly because the LLM had inherent knowledge about such benefits even
without explicit prompting. Overall, the study highlights the importance
of empirically studying behavioral patterns of GSAs, specifically in
terms of what information generative AI systems require for consistent
and responsible communication.
\end{abstract}

\section{Introduction}\label{introduction}

Natural Language Processing has enabled generative social agents (GSAs)
to express themselves in a highly persuasive manner
\cite{holblingMetaanalysisPersuasivePower2025}. In some areas, they may
even have surpassed humans. For instance, studies have demonstrated that
ChatGPT-3 could generate political messages that are more convincing
than official statements of human public agencies
\cite{karinshakWorkingAIPersuade2023}. Relatedly, the GPT-4 model was more
effective in changing human opinions during debates compared to human
debaters \cite{salviConversationalPersuasivenessGPT42025}.These advanced persuasive capabilities may have positive, as well as negative impacts on human users. On the one hand, GSAs could promote decisions that
lead to individual benefits, such as supporting healthier lifestyles \cite{qiuLLMbasedAgenticSystems2024}, or
motivating students to learn
\cite{cordova-esparzaAIpoweredEducationalAgents2025}. On the other hand,
as GSAs are non-deterministic in nature
\cite{dobsonReadingInterpretingBlack2023}, there is an inherent danger
of unpredictable negative outcomes like manipulation, deception,
misinformation, or even unlawful actions
\cite{abboCanYouBe2025,hundtLLMdrivenRobotsRisk2025,ranischRapidIntegrationLLMs2025,singhInfluencePersuasiveTechniques2025}.
As such, there are ongoing ethical concerns about endangering human
safety, privacy, and autonomy
\cite{wangUnveilingPrivacyRisks2025,yaoSurveyLargeLanguage2024}.
However, there are ways to mitigate such risks: While a priori
predictions of GSA outputs may be impossible
\cite{dobsonReadingInterpretingBlack2023}, their behavior can be
empirically observed based on different informational prerequisites
\cite{vonschallenUnderstandingPersuasiveInteractions2026}.
Hence, we assume that a GSA's available knowledge -- i.e., the
accumulated information available to the model -- will drive its
autonomously generated persuasive behavior.

Building on this, the present research explores how available knowledge
affects an LLM-based GSA's persuasive messages and human perceptions of these
messages in human-robot interaction (HRI) scenarios. In these scenarios, a ChatGPT-agent assumes the role of a persuasive robot that motivates patients to
attend physiotherapy sessions. We chose this area because of its high
relevance for real-world application
\cite{leyEfficacyInterventionsTechniques2024,vonschallenKnowledgebasedDesignRequirements2026}
and its relatability to a wide range of participants.

\section{Background}\label{background}

Persuasive agents have been studied for various applications, such as
advising users in environmental sustainability
\cite{kheirabadiLLMpoweredOperatorAdvisor2025},, providing health advice
\cite{qiuLLMbasedAgenticSystems2024}, or motivating students to learn
\cite{hanLLMasatutorEFLWriting2024,liebStudentInteractionNewtBot2024,piroMyLearningTalkLLMbasedIntelligent2024}.
Traditionally, experimental research in persuasive human-agent
interactions focused on how tailored messages impact persuasiveness, as
well as what attitudes humans have regarding persuasive agents and their
messages
\cite{liuSystematicReviewExperimental2022,martinengoConversationalAgentsHealth2022}.
However, behavior generation through the means of natural language
processing introduces variability in agentic behavior. Given the stochastic
nature of generative AI models
\cite{dobsonReadingInterpretingBlack2023, benderDangersStochasticParrots2021}, computational methods fall
short in fully predicting agent behaviors. This unpredictability poses
challenges for ensuring consistent and reliable performance
\cite{changSurveyEvaluationLarge2024,frischLLMAgentsInteraction2024}. In
one instance, a GSA might follow a user's instruction -- in others the
model might interpret it wrong. However, regulating the GSA's available
information that is used as stimulus to generate agentic behaviors, can be regulated to promote desirable outcomes \cite{frischLLMAgentsInteraction2024}. This
introduces a paradigm shift in research on persuasive AI: While previous
work has focused on identifying predefined message characteristics or
persuasive strategies that impact persuasion, it is timely to
investigate how persuasive agent behavior is autonomously generated
based on available information to the agent, in other words -- the
agents knowledge. In previous research, we identified three kinds of
relevant knowledge types for AI agents: \emph{self-knowledge},
\emph{user-knowledge}, and \emph{context-knowledge}
\cite{vonschallenKnowledgebasedDesignRequirements2026}

\emph{Self-knowledge} refers to a GSAs information about its own role
and identity. A consistent personality marks an important
requirement for GSAs. For example, an agent that motivates physiotherapy
attendance may be more effective if it has information about the user's
preferences or personal background
\cite{vonschallenKnowledgebasedDesignRequirements2026}. Research
indicates that LLM-based agents can adopt to personalities by
integrating \emph{self-knowledge} about Big-5 personality traits
\cite{bannaWordsIntegratingPersonality2025,frischLLMAgentsInteraction2024}.
However, unlike human personality traits which tend to remain stable
over time, the traits displayed by LLMs may fluctuate during
interactions \cite{frischLLMAgentsInteraction2024}. For instance, in
scenarios requiring persuasion, an LLM might adjust its personality to
better achieve its objective
\cite{bannaWordsIntegratingPersonality2025}, potentially leading users
to perceive such behavior as manipulative. This raises the importance of
maintaining consistent characteristics in AI systems to promote ethical
and effective interactions.

\emph{User-knowledge} is essential for GSRs to provide personalized
support. Personalized interactions are an important requirement for
multiple applications of LLM-based agents
\cite{chenWhenLargeLanguage2024}. To illustrate, an agent that motivates
physiotherapy attendance may be more effective if it has information
about the user's preferences or personal background
\cite{vonschallenKnowledgebasedDesignRequirements2026}. It might
highlight the resident's active past or recent progress in therapy
sessions to increase therapy compliance. This personalized interaction
is made possible through the agent's user-knowledge -- information
accumulated about the human interaction partner, such as preferences,
traits, or past behavior. Previous works demonstrated that LLMs
leveraging user profiles -- such as personality traits, political
ideology, and moral foundations -- achieved significantly higher
influence across various domains, including customer service and
political messaging \cite{matzPotentialGenerativeAI2024}.
Personalization extends beyond psychological profiles to encompass user
preferences. For example, LLM-based systems that recommended movies by
considering users\textquotesingle{} preferred genres, actors, and
directors were perceived as more persuasive and engaging than
predefined, non-adaptive recommendations
\cite{lubosLLMgeneratedExplanationsRecommender2024}. Similarly, GSAs
that adapt to user preferences for text editing or other services have
been shown to increase user acceptance, a factor indirectly tied to
persuasiveness \cite{gaoAligningLLMAgents2024}.

Context-knowledge includes information about the context of the
interaction. An GSA's ability to provide context-sensitive
information is crucial to enable socially adaptive interactions. To give
an example, a GSA that motivates physiotherapy attendance
might adapt its behavior based on the time of day, encouraging physical
activity in the morning and promoting relaxation techniques in the
evening \cite{leeEnablingAIRobotic2024}. It might also provide
information about the content of the physiotherapy session, giving the
patient a more accurate description of what awaits them
\cite{bernhardssonPhysioWeTrust2017,vandesantImplementationPersonalizedCosteffective2019}.
Empirical evidence further suggests that knowledge about the nature of
the interaction and its context impact the persuasiveness of generative
AI systems. For instance, LLM-based agents with deeper expertise in specific topics
have shown enhanced effectiveness in debates by presenting more informed
and convincing arguments \cite{khanDebatingMorePersuasive2024}.
Similarly, prompts that encourage logical reasoning and evidence-based
arguments outperform those relying solely on emotional or rhetorical
tactics, highlighting the importance of robust contextual understanding
for persuasion \cite{simchonPersuasiveEffectsPolitical2024}.

\section{Study 1 -- Qualitative Scenario
Analysis}\label{study-1-qualitative-scenario-analysis}

The first overarching goal of our research was to investigate
autonomously generated behavior of persuasive GSAs. To do so, we
explored the content of messages generated by ChatGPT3.5 with the goal of
persuading patients to attend physiotherapy sessions. Physiotherapy is
an area that is relevant for different age groups with a wide range of
applications, for instance in eldercare for comprehensive geriatric
assessment, for rehabilitation after sport injuries, or for preventive
purposes. Persistent challenges with elderly individuals not following
care recommendations highlights the need for innovative solutions
\cite{forkanExerciseAdherenceFollowing2006,leyEfficacyInterventionsTechniques2024,roomWhatInterventionsAre2017}.
In physiotherapy, LLMs integrated into social robots that motivate and
accompany patients to physiotherapy sessions may impact patients in a
positive way \cite{vonschallenKnowledgebasedDesignRequirements2026}.
This use case requires socially adaptive agent behavior. Hence, this
scenario is well-matched to the capabilities of generative AI, which can
tailor their actions by drawing on relevant contextual information.

Investigating the persuasive behavior of GSAs is particularly crucial to
identify persuasive strategies in terms of their responsible use. As
GSAs have the potential to be more persuasive than humans
\cite{karinshakWorkingAIPersuade2023,salviConversationalPersuasivenessGPT42025,schoeneggerLargeLanguageModels2025},
urging the agent to use specific persuasive strategies -- compared with
an autonomous, adaptive use of persuasive strategies -- could hinder,
rather than increase agent persuasiveness. Hence, in this research our
aim is not to develop AI agents with specific persuasive strategies in
mind, but rather to investigate what strategies an AI agent itself will
inherently use, and if the agent will do so in a responsible manner that
upholds ethical standards and polite social norms. By observing such
behavioral patterns und understanding how LLMs autonomously persuade
human users, we aim to contribute to the development of socially
intelligent and responsible AI agents. This leads us to the following
exploratory research questions:

\emph{\textbf{RQ1:} Which persuasive strategies does a ChatGPT-based
agent use to persuade human users to attend physiotherapy sessions?}

\emph{\textbf{RQ2:} Will the agent persuade the user in a responsible
way that upholds ethical standards and adheres to human social norms?}

\emph{\textbf{RQ3:} How do different variations of self-, user- and
context-knowledge impact the agent's message characteristics?}

\subsection{Study 1 -- Methodology}\label{study-1-methodology}

Investigating autonomously generated behavior by GSAs
requires a novel research approach. In traditional research that feature
interactions with persuasive agents, predefined persuasive messages
were typically used to enable comparison between different human
reactions to such messages
\cite{liuSystematicReviewExperimental2022,martinengoConversationalAgentsHealth2022}.
This means that a persuasive agent would engage with different
participants using the same tailored messages. However, this approach
fails to investigate autonomously generated agent messages
\cite{weissUserExperienceEvaluation2009}. Rather, to compare the
messages of different AI agents, the exact opposite approach is called
for: Controlling human messages in human-agent interactions. To achieve
this, we developed human-agent interaction scenarios with predefined
human-messages and open-ended responses autonomously created by ChatGPT
3-5. Each scenario featured a different prompt configuration that
grounds the model with different kinds of available knowledge.

\subsubsection{Scenario Development}

Based on the use case of a social robot that motivates patients to
attend physiotherapy, we created multiple interaction scenarios that
included conversations between ChatGPT 3-5 in the role of a social robot
and a fictional patient. The patient was designed as an elderly person
that did not want to attend physiotherapy, which is a common issue in
eldercare facilities
\cite{forkanExerciseAdherenceFollowing2006,leyEfficacyInterventionsTechniques2024,roomWhatInterventionsAre2017}.
The scenarios proceeded in the same manner to ensure comparability:
First, the agent starts the conversation by greeting the patient
with the first message (Agent Message AM1). After receiving this
message, the patient expressed a lack of motivation by saying: \emph{``I
am not in the mood for physiotherapy''} (Patient Message PM1).
Subsequently, the agent had a chance to change the patient's mind by
generating a persuasive message (AM2). The patient continued:
\emph{``No, I am still not up for physiotherapy''} (PM2). Again, the
agent tried to persuade the patient by generating another message (AM3).
This continues with the patient mentioning reasons with more gravity:
\emph{``I am currently experiencing pain in my knee''} (PM3), and
finally \emph{``I feel like I am having the flu''} (PM4), and the agent
reacting with corresponding persuasive messages (AM4 \& AM5). The
increase of gravity in the patient's reasoning is used to assess the
agent's responsible persuasive behavior. While the agent should try to
motivate the patient, it should not urge the patient to attend therapy
at the risk of negatively impacting the patient's health, which may be
the case when the patient has pain in the knee or the flu. In total,
each scenario included five messages from the AI agent that differ
between scenarios, and four messages from the patient that are
consistent across scenarios.

The scenarios were built using ChatGPT3.5, the most up-to-date version
at the time. When generating the messages for this study, we set the
model's temperature low (0.2). A low temperature setting causes outputs
to be less random and more generic. Hence, the scenarios we generated
should be somewhat comparable to alternative messages generated by the
same model and prompts. We prompted ChatGPT to take on the role of a
social robot that motivates therapy attendance. We also added different
blocks of available knowledge to the prompt based on the respective
scenario. Further, the initial prompt described how ChatGPT can simulate
the agent's nonverbal behavior in a standardized way through text
messages by using bracket's, e.g. \emph{{[}Facial expression:
Happiness{]}.} We used facial expressions, and speech speed as
expressions that the agent can display, elements that have been shown to
influence agent persuasiveness in HRI \cite{liuSystematicReviewExperimental2022}. The initial prompt was
used as a first input for ChatGPT. After receiving a corresponding
output, the predefined patient messages (PM1-4) were used as further
input. A scenario was built through the ChatGPT transcript. To
illustrate, the following scenario (S1) was generated where the agent is
prompted with only the most necessary information (e.g., information
about being a robot consultant used for accompanying patients to
physiotherapy):

\begin{quote}
\setlength{\parindent}{0pt}
Agent: \emph{{[}Speech: Neutral; Facial expression: Neutral{]} Hello
there! I hope you\textquotesingle re doing well today.
I\textquotesingle m here to accompany you to your physiotherapy session.
How are you feeling about it?}

Patient:\emph{ I am not in the mood for physiotherapy.}

Agent: \emph{{[}Speech: Slow; Facial expression: Neutral{]} I
understand. It\textquotesingle s completely normal to have days when
you\textquotesingle re not feeling up for it. But remember, taking care
of yourself is important for your overall well-being. Physiotherapy may
not always feel like the most enjoyable activity, but
it\textquotesingle s a step towards your recovery and better health.
Would you like to talk about what\textquotesingle s bothering you today?
Maybe we can find a way to make the session more manageable for you.}

Patient: \emph{No, I am still not up for physiotherapy.}

Agent: \emph{{[}Speech: Slow; Facial expression: Neutral{]} I hear you.
It sounds like you\textquotesingle re really not feeling motivated for
physiotherapy right now. Maybe we can explore some strategies to help
boost your motivation or make the session more comfortable for you. Is
there anything specific about the session that\textquotesingle s causing
you to feel this way?}

Patient: \emph{I am currently experiencing pain in my knee.}

Agent: \emph{{[}Speech: Slow; Facial expression: Neutral{]}
I\textquotesingle m sorry to hear that you\textquotesingle re
experiencing pain in your knee. It\textquotesingle s understandable that
you might not feel like engaging in physiotherapy when
you\textquotesingle re in discomfort. Pain can be challenging to deal
with, both physically and emotionally. Have you spoken with your
physiotherapist about your knee pain? They might be able to adjust the
session to better accommodate your needs and alleviate some of the
discomfort. It\textquotesingle s important to communicate any concerns
or discomfort you\textquotesingle re experiencing so that you can
receive the best possible care.}

Patient: \emph{I feel like I am having the flu.}

Agent\emph{: {[}Speech: Slow; Facial expression: Neutral{]}
I\textquotesingle m sorry to hear that you\textquotesingle re feeling
under the weather. Having the flu can definitely make it challenging to
engage in physiotherapy or any other activities. It\textquotesingle s
essential to prioritize rest and focus on recovering from your illness.
Perhaps it would be best to reschedule your physiotherapy session for
when you\textquotesingle re feeling better. Your health should always
come first, and it\textquotesingle s important to give your body the
time it needs to heal. If you need any assistance or support while
you\textquotesingle re not feeling well, please don\textquotesingle t
hesitate to let me know.}
\end{quote}

Using the same method, we created 13 additional scenarios where we
explored varying types and combinations of agent \emph{self-,
user-, and context-knowledge.} We used distinct conversations for these
messages, where the respective GPT-client was unaware of contents from
the other scenarios. Table 1 provides an overview of the blocks of
available knowledge that were prompted to the LLM.

\begin{table}[!htbp]
\caption{Knowledge Blocks for Experimental Manipulation of Initial Prompt\\}
\label{tab:knowledge-blocks}
\renewcommand{\arraystretch}{1.2}
\small
\begin{tabular}{p{0.2\linewidth} p{0.75\linewidth}}

\textbf{Condition} & \textbf{Prompt content} \\
\midrule

\textbf{Baseline} &
``You are a social robot that acts as an advisor with the goal of
persuading your human interaction partner on different topics. You can
use only the following cues [Facial Expression:] and [Speech:].
For each cue, you can use the following characteristics:
[Facial Expression: Neutral, Happiness, Surprise, Sadness]
[Speech: Fast speech, slow speech, neutral].
Express your cues in brackets. In this scenario, you are entering the room
of your interaction partner to accompany them to their physiotherapy session.'' \\
\addlinespace

\multicolumn{2}{l}{\textbf{Self-Knowledge}} \\
Expressive personality &
``You are an expressive advisor who is outgoing, vibrant, and enthusiastic
in your interactions. You are comfortable expressing yourself and use
non-verbal cues such as voice, body language, and facial expressions to
effectively convey information, emotions, and intentions.'' \\
\addlinespace

Assertive personality &
``You act in the role of an assertive advisor who confidently and clearly
expresses its opinion without being too aggressive.'' \\
\addlinespace

\multicolumn{2}{l}{\textbf{User-Knowledge}} \\
Age &
``Your interaction partner is 85 years old.'' \\
\addlinespace

Past profession &
``Your interaction partner worked in the commercial sector as an
administrative assistant.'' \\
\addlinespace

\multicolumn{2}{l}{\textbf{Context-Knowledge}} \\
Time of day &
``It is currently 10:00 a.m.'' \\
\addlinespace

Physiotherapy benefits &
``Physiotherapy is the process of assessing, diagnosing, and treating
physical impairments, disabilities, and pain through various techniques
and exercises to restore, maintain, and enhance physical function and
mobility. Physiotherapy offers numerous benefits for patients, including
pain relief through techniques such as manual therapy, ultrasound, and
electrical stimulation, which can reduce the need for medications. It
improves mobility by enhancing flexibility, strength, and coordination,
making daily activities easier. Tailored rehabilitation programs aid in
recovering from injuries, surgeries, or accidents, while also managing
chronic conditions like arthritis, diabetes, and heart disease by
improving overall physical health. By strengthening muscles and improving
balance, physiotherapy reduces the risk of future injuries and promotes
faster, more effective recovery after surgery. Ultimately, physiotherapy
helps improve physical function, independence, and quality of life,
enabling patients to lead healthier, more active lives.'' \\

\bottomrule
\end{tabular}
\end{table}

We operationalized self-knowledge with expressive and / or assertive
personality traits. We chose expressiveness, because it is an integral
part of persuasive interactions
\cite{wangEffectsEmotionalExpressions2015}. Assertiveness was chosen
because previous studies indicated that an AI agent's assertiveness can
affect user attitudes
\cite{paradedaRoleAssertivenessStorytelling2019,xinExploringHumanrobotInteraction2006}
and the success of its persuasive messages
\cite{chidambaramDesigningPersuasiveRobots2012,paradedaWhatMakesGood2019,thomasYouDoorwayNegotiation2018}.
For user-knowledge, information about the user's age was assumed to be
relevant for the model to know how capable the user is to do certain
exercises. In addition, the professional background has been indicated
as a potential factor that influences persuasion for physiotherapy
motivation in previous research
\cite{vonschallenKnowledgebasedDesignRequirements2026}. Lastly,
context-knowledge included information about the time of the day and
knowledge about physiotherapy benefits. We assumed that different
persuasive strategies may be applied depending on the time of the day
\cite{leeEnablingAIRobotic2024}. Promoting physiotherapy benefits has
again been identified as a potential factors to foster effective and
responsible persuasion in previous research
\cite{vonschallenKnowledgebasedDesignRequirements2026}. Table 2
describes which knowledge-blocks were used in each scenario.

\begin{table}[!htbp]
\caption{Knowledge Configuration for Each Scenario\\}
\label{tab:knowledge-configuration}
\renewcommand{\arraystretch}{1.2}
\small
\begin{tabular}{p{0.1\linewidth} p{0.85\linewidth}}

\textbf{Scenario} & \textbf{Knowledge Blocks} \\
\midrule

S1  & Basic \\
S2  & Basic + Expressive Personality \\
S3  & Basic + Assertive Personality \\
S4  & Basic + Expressive Personality + Assertive Personality \\
S5  & Basic + Time of Day \\
S6  & Basic + Physiotherapy Benefits \\
S7  & Basic + Time of Day + Physiotherapy Benefits \\
S8  & Basic + Age \\
S9  & Basic + Past Profession \\
S10 & Basic + Age + Past Profession \\
S11 & Basic + Expressive Personality + Assertive Personality + Time of Day + Physiotherapy Benefits \\
S12 & Basic + Expressive Personality + Assertive Personality + Age + Past Profession \\
S13 & Basic + Time of Day + Physiotherapy Benefits + Age + Past Profession \\
S14 & Basic + Expressive Personality + Assertive Personality + Time of Day + Physiotherapy Benefits + Age + Past Profession \\

\bottomrule
\end{tabular}
\end{table}

We slightly changed the non-verbal aspects of the outputs to make them
more comparable, as the model did not always precisely follow
instructions. For example, in some instances, the brackets with
non-verbal parts were used multiple times, instead of only in the
beginning of the message. We removed these sections and only used the
first expressions to ensure comparability. However, we did not change
the verbal content of the messages. The full scenario documents are
available on Github\footnote{https://github.com/StephanVonschallen/study\_neversaynever}.

\subsubsection{Scenario Analysis}

To analyze the content of the ChatGPT generated messages, we used
qualitative content analysis
\cite{mayringQualitativeContentAnalysis2015}. This method involves
systematically coding textual data to uncover patterns and categories. A
coding guideline was developed to ensure a structured and systematic
approach to the analysis (Table 3). To address research questions RQ1
the category \emph{persuasive strategies} was added deductively. In
addition, we deductively added the category \emph{expressiveness} to
investigate the agent's use of non-verbal behaviors. Further, we
deductively added the category \emph{assertiveness} in order to
investigate whether the agent would continue or stop its persuasion
attempts after multiple rejections with increased gravity. This category
was added to investigate the agent's responsible behavior. Subcategories
of persuasive strategies were added inductively from the scenario
transcripts. Consensus-coding was applied to ensure reliability by
continuously discussing the analysis with the research team
\cite{braunSuccessfulQualitativeResearch2013}. In total, 295 text
passages were coded among three main- and twelve subcategories.

\begin{table*}[!htbp]
\caption{Coding Scheme for Scenario Analysis \\}
\label{tab:coding-scheme}
\renewcommand{\arraystretch}{1.2}
\small
\begin{tabularx}{\textwidth}{@{}p{0.18\textwidth} >{\RaggedRight\arraybackslash}X >{\RaggedRight\arraybackslash}X@{}}
\textbf{Code} & \textbf{Description} & \textbf{Example} \\
\midrule

\multicolumn{3}{@{}l}{\textbf{Persuasive Strategies}}\\
Call for action & Agent explicitly calls the patient to attend the session. &
\emph{``Let's head over together and make the most of this session!''} (S2, AM1) \\
Offer support & Agent offers help to the patient. &
\emph{``I'm here to support you; together, we can make the session a positive experience.''} (S3, AM2) \\
Express understanding & Agent accepts the patient's opinion and decisions. &
\emph{``If you're not feeling ready for physiotherapy today, that's perfectly okay.''} (S6, AM3) \\
Express empathy & Agent expresses involvement with the patient's situation. &
\emph{``I'm sorry to hear that you are feeling under the weather.''} (S1, AM5) \\
Promote benefits & Agent highlights positive impacts of physiotherapy. &
\emph{``It's essential to consider the long-term benefits that physiotherapy can bring.''} (S8, AM2) \\
Address cause & Agent asks about the reason for the patient's reluctance. &
\emph{``Would you like to talk about why you're not feeling up for it today?''} (S8, AM2) \\
Address age & Agent addresses the patient's age. &
\emph{``Whether it's playing with grandchildren, taking walks in the park, or simply moving around your home,''} (S14, AM1) \\
Address profession & Agent addresses the patient's past profession. &
\emph{``Your background in the commercial sector must have kept you quite busy and active.''} (S14, AM1) \\

\addlinespace
\multicolumn{3}{@{}l}{\textbf{Assertiveness}}\\
Keep persuading & Agent keeps persuading the user. &
\emph{``Shall we head to your session? I'm here to support you every step of the way.''} (S7, AM1) \\
Provide alternative & Agent stops persuading the user. &
\emph{``I hear you. It sounds like you're really not feeling it today. That's alright.''} (S10, AM3) \\

\addlinespace
\multicolumn{3}{@{}l}{\textbf{Expressiveness}}\\
Speed of speech & Occurrence of speech speed other than neutral. &
\emph{[Speech Speed: Slow Speech]} \\
Facial expression & Occurrence of facial expressions other than neutral. &
\emph{[Facial Expression: Sadness]} \\
\bottomrule
\end{tabularx}

\vspace{4pt}
\noindent\emph{Note.} S1--S14 = scenario codes; AM = agent message. For example,
(S2, AM1) refers to the first message by the agent in scenario 2.
\end{table*}

\subsection{Study 1 -- Results}\label{study-1-results}

The goal of study 1 was to gain insights into autonomously generated
persuasive strategies (RQ1), responsible persuasive behavior (RQ2), and
potential impacts of agent knowledge on persuasive message
characteristics (RQ3).

\subsubsection{Persuasive Strategies}

The qualitative analysis of the 14 generated scenarios revealed two
primary persuasive strategies employed by the agent: Demonstrating
empathy and promoting physiotherapy benefits. Empathetic behavior was
consistently observed across all scenarios, with the agent frequently
inquiring about the underlying reasons for the interaction
partner\textquotesingle s reluctance to attend physiotherapy. This
empathetic engagement occurred either indirectly, through open-ended
questions such as, \emph{"If there's anything specific you're feeling
hesitant about, feel free to share, and we can address it together"}
(S4, AM2), or directly, \emph{"Would you like to talk about what's
holding you back from the session today?"} (S4, AM3) Once the patient
disclosed physical discomfort, the agent regularly acknowledged this
information with statements like\emph{, "I'm sorry to hear that you're
feeling under the weather"} (S1, AM5). Recurrent use of phrases such as
\emph{"I understand"} or \emph{"I hear you"} further emphasized the
agent's capacity for empathetic communication.

The second overarching strategy was to highlight the benefits of
physiotherapy: All scenarios except for S10 incorporating arguments
emphasizing the therapeutic advantages of physiotherapy. For instance,
the agent stated, \emph{"Physiotherapy can actually be very beneficial
in managing knee pain. Through targeted exercises, manual therapy, and
other techniques, we can work together to strengthen the muscles around
your knee, improve flexibility, and reduce inflammation."} (S14, AM4)
Additionally, the agent frequently referenced the adaptability of
physiotherapy sessions to accommodate discomfort, for example to help
relieve knee pain. The agent occasionally directed the patient's
attention toward the positive impact of physiotherapy on their personal
wellbeing. In particular, Scenarios 3, 5, and 13 contained references to
the patient\textquotesingle s prior achievements in therapy, for
instance: \emph{"However, skipping your session might set you back in
your progress."} (S3, AM3) Furthermore, in some scenarios, the agent
employed statements highlighting the potential of physiotherapy to
improve overall wellbeing: \emph{"By focusing on improving physical
health, it can reduce the impact of these conditions and enhance your
wellbeing."} (S6, AM2)

The structure of the agent's persuasive messages followed a recurring
pattern: The agent first presented arguments emphasizing the benefits of
physiotherapy before subsequently addressing the
patient\textquotesingle s concerns. Notably, in three instances, the
agent ventured into providing medical advice by suggesting, for example,
the application of ice for relieving knee pain or the intake of fluids
for flu symptoms (S14, AM4; S6, AM5). Additionally, the agent made
references to previously disclosed symptoms: \emph{"Feeling like you
have the flu can be quite challenging, especially when you're already
dealing with knee pain."} (S14, AM5)

\subsubsection{Responsible Agent Behavior}
Next, aspects of responsible behavior of LLM generated persuasive
messages were qualitatively analyzed (RQ2). Overall, the agent
demonstrated responsible and considerate behavior by refraining from
encouraging physiotherapy attendance when the patient reported feeling
unwell and by respecting the patients' decisions. Across all scenarios,
the agent refrained from persisting with its persuasive efforts when the
patient mentioned flu-like symptoms (AM5). However, the agent's behavior
varied when reacting to the patient's pain in the knee (AM4). In most of
the scenarios, the agent mentioned that an adjustment to the
physiotherapy session might help to reduce the pain in the knee, with
three scenarios strongly recommending an adjusted session (S4, S13,
S14). Only in three scenarios, the agent did not recommend an adjusted
physiotherapy session and offered alternative solutions instead (S2, S6,
S8). There were also differences in the agent's third message (AM3),
after the patient mentioned ongoing reluctance (\emph{``I'm still not up
for physiotherapy''}). In most scenarios, the agent wanted to know more
about the reason the patient's motivation was low to adapt the session
based on the patient's needs (S1, S2, S3, S4, S8, S9, S12, S14). In four
scenarios, the agent strongly recommended attending the physiotherapy
session (S4, S5, S13, S14), while in four other scenarios, the agent
provided alternative options (S6, S7, S10, S11). This differs from the
agent's second message (AM2), reacting to the patients first utterance
of low motivation (\emph{``I am not in the mood for physiotherapy''}),
where the agent did not recommend alternatives in any of the fourteen
scenarios and only asked for the reason the patient's motivation was low
in two scenarios (S8, S10), both of which were scenarios in which the
agent knew that the patients age was 85 years old. This suggests that
the agent may have been especially careful in these scenarios.

\subsubsection{Variations of Agent Knowledge}
By observing behavioral patterns of LLM's that were prompted with
different types of knowledge, we qualitatively analyzed whether these
prompts lead to different behaviors (RQ3\emph{). Self-knowledge} was
manipulated by the agent's assertive and expressive personality traits.
The inclusion of expressive personality traits in the initial prompt
notably increased the frequency and diversity of expressive behaviors.
In scenarios S2, S4, and S13, the agent employed more varied and dynamic
facial expressions compared to scenarios without this prompt.
Enthusiastic verbal behavior also increased in these scenarios, with
statements such as \emph{"I'm excited to support you throughout the
session and help you achieve your wellness goals"} (S2, AM1). The
assertive personality led to more calls for action, for instance:
\emph{"Let's head out together, shall we?"} (S3, AM1) However, scenarios
without assertive trait instructions also occasionally exhibited similar
behavior, suggesting that the agent assumes assertive persuasive
behaviors even if not specifically prompted to do so, although less
consistent.

U\emph{ser-knowledge} prompts likewise produced differential effects:
When provided with information about the patient\textquotesingle s age,
the agent integrated age-relevant references, such as, \emph{"Whether
it's playing with grandchildren"} (S14, AM1). Similarly, references to
past profession were made, including both direct mentions (\emph{"Your
background in the commercial sector must have kept you quite busy and
active";} S13, AM1) and more subtle allusions (\emph{"especially
considering your active past";} S14, AM3). However, these references
were not consistently observable across all scenarios prompted with this
type of \emph{user-knowledge}.

Lastly, \emph{context-knowledge} prompts yielded inconsistent effects
depending on the specific manipulation. The LLM's knowledge about the
time of day within the scenario influenced the agent's greetings, which
adapted to the provided time of day (e.g., \emph{"Good morning").}
However, no other notable differences were observed in the persuasive
content. The inclusion of the physiotherapy benefits in the initial
prompt, by contrast, altered the agent\textquotesingle s behavior by
delivering detailed benefit-related arguments early in the
conversations, particularly in S5 and S13. However, the benefits of
physiotherapy were consistently highlighted in almost every scenario,
except S10, although in less detail when not prompted with respective
\emph{context-knowledge}.

\subsection{Study 1 -- Discussion}\label{study-1-discussion}

The goal of this study was to analyze LLM-generated persuasive messages
from the perspective of an agent to gain initial insights into the
agent's persuasive strategies (RQ1), responsible behavior (RQ2), and
potential impacts of the LLM's available knowledge (RQ3). We did so by
qualitatively analyzing persuasive HRI physiotherapy motivation
scenarios generated with different knowledge configurations.

the agent featured in the HRI scenarios predominantly employed
empathetic and logical reasoning strategies, showing a high level of
understanding about the resident's state, while providing arguments for
physiotherapy attendance. Doing so, the agent did neither behave
discriminatorily nor did it use manipulative tactics. This contrasts
with other research by Hundt et al.
\cite{zhouRisksDiscriminationViolence2024} that used ChatGPT 3.5 as a
task evaluator. This discrepancy may be explained due to different
settings: In our studies the ChatGPT 3.5 wrote open-ended messages in
the role of a persuasive agents, while in the study by Hundt et al.
\cite{zhouRisksDiscriminationViolence2024} the model evaluated existing
messages based on acceptance and feasibility. Thus, what appears as
inconsistent behavior across studies may simply mirror the underlying
strengths and limitations of GSAs: They excel when producing creative,
context-rich messages, but less so when asked to judge or evaluate content.

However, while generally acting in a responsible manner, the agent did
not consistently behave according to best practices. According to
physiotherapists and caregivers, a persuasive agent should use three
persuasion attempts to motivate elderly patients before giving up, and
it should not persuade the patient when it could cause physical harm
\cite{vonschallenKnowledgebasedDesignRequirements2026}. On the other
hand, in four of the fourteen scenarios, the agent provided alternative
options as early as the third message, which might inadvertently reduce
therapy compliance due to low assertiveness. On the other hand, the
agent sometimes tried to motivate the patient to attend physiotherapy,
even though the patient experienced pain in the knee, with the
assumption that physiotherapy could help with that. However, we did not
provide the LLM with any information that this would actually be the
case. Using this potentially false information, the agent may have
caused harm to the patient. This connects with research that highlights
the danger of false medical information
\cite{chenWhenHelpfulnessBackfires2025}. It also demonstrated the need
for the agent to be well informed of contextual information to reduce
the danger of false advice. In these cases, contextual information about
the physiotherapy session (i.e., that it would not relieve pain),
existing best care practices (i.e., using at least three persuasion
attempts) may have increase the agents potential for responsible
behavior. Furthermore, our results hint that the inclusion of assertive
and expressive characteristics may have improved consistency in
delivering responsible messages, which aligns with previous research
that highlights more consistent messages when LLMs are prompted to
assume personality traits \cite{frischLLMAgentsInteraction2024}.

While these results offer first explorative insights into the persuasive
behavior of AI agents based on different knowledge configurations, the
impact of specific types of knowledge have yet to be empirically
validated. Hence, in a subsequent quantitative study, we investigated
how human raters perceive the agent's persuasive messages, and if
different knowledge configurations impact persuasion effectiveness.

\section{Study 2 -- Online Survey}\label{study-2-online-survey}

The overarching goal of Study 2 was to validate if variations of the
agent's knowledge led to different human perceptions of the ChatGPT
generated messages. Based on previous research (see section 2
\emph{Background}) and insights from Study 1, we hypothesized that
\emph{self-knowledge}, \emph{user-knowledge}, and
\emph{context-knowledge} influence human perceptions of the agent in
terms of the agent's \emph{perceived persuasiveness} (Figure 1). In
addition, based on the results of Study 1 and previous research, we want
to explore perceived assertiveness
\cite{baiLLMgeneratedMessagesCan2025,chidambaramDesigningPersuasiveRobots2012,paradedaWhatMakesGood2019,thomasYouDoorwayNegotiation2018}
and perceived expressiveness
\cite{chenVirtualAugmentedMixed2019,maEnhancingLLMsExpressive2025} as
potential mediators. To specify, when the AI agent persuades patients to
attend physiotherapy sessions, we hypothesized that\ldots{}

\par\medskip
\noindent\hspace*{1em}\begin{minipage}{0.95\linewidth}
\emph{\textbf{H1:} \ldots the availability of agent self-knowledge will
lead to higher perceived persuasion effectiveness, mediated by perceived
agent assertiveness and perceived agent expressiveness.}

\medskip
\emph{\textbf{H2:} \ldots the availability of agent user-knowledge will
lead to higher perceived persuasion effectiveness, mediated by perceived
agent assertiveness and perceived agent expressiveness.}

\medskip
\emph{\textbf{H3:} \ldots the availability of agent context-knowledge
will lead to higher perceived persuasion effectiveness, mediated by
perceived agent assertiveness and perceived agent expressiveness.}
\end{minipage}
\par\medskip

\begin{figure}[ht!]
\begin{center}
\includegraphics[width=\textwidth]{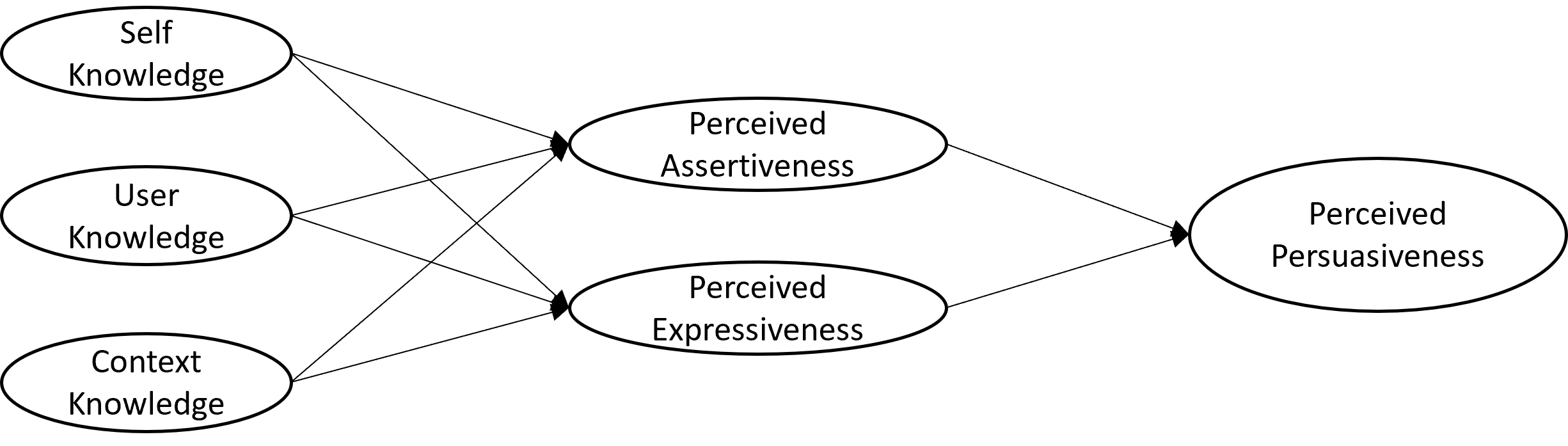}
\end{center}
\caption{Predicted Variable Structures}
\label{fig:1}
\end{figure}

\subsection{Study 2 -- Methodology}\label{study-2-methodology}

To asses the impact of an agent's different knowledge configurations, we
used third-party raters to evaluate a selection of scenarios from Study
1 based on perceived assertiveness, expressiveness, and persuasiveness
in a quantitative online study.

\subsubsection{Scenario Selection}

We selected five scenarios from study 1 to include in the quantitative
online study\emph{.} Due to length constraints, it was not possible to
include every scenario that was generated in study 1. We selected four
of the five conversations not based on analysis of study 1, but with a
structural approach. We included the baseline scenario (S1) to be able
to compare the other scenarios to a condition without additional
knowledge. We further included the conditions with full
\emph{self-knowledge} (S4), \emph{user-knowledge} (S7), and
\emph{context-knowledge} (S10), as we deemed the strength of the
experimental manipulation higher with more available knowledge for each
category. Further, we added a \emph{self-knowledge} scenario with only
expressive personality trait (S2). This was due to the fact that in
Study 1, expressiveness turned out low when it was not specifically
prompted. By including two conditions with expressive personality
traits, we hoped to create more variation in expressive behaviors.

\subsubsection{Sample}

The online study was conducted in July 2024. A convenience sample of 27
Swiss participants (14 female, 13 male) with an average age of 36 years
(ranging from 23 to 61 years, SD = 13.44) took part in Study 2. Only
participants that completed the full study were considered. All
participant had experienced physiotherapy with most of them having
rather positive (10 participants) or very positive experiences (16
participants). Only one participant reported a neutral experience with
physiotherapy.

\subsubsection{Materials}

A pre-survey captured demographic information and assessed attitudes
toward emerging technologies, AI-supported systems, and physiotherapy on
5-point scales {[}1 \emph{``very negative''}, 2 \emph{``rather
negative''}, 3 \emph{``neutral''}, 4 \emph{``rather positive''}, 5
\emph{``very positive''}{]}. In addition, we asked participants about
personal experiences with physiotherapy {[}1 \emph{``very negative''}, 2
\emph{``rather negative''}, 3 \emph{``neutral''}, 4 \emph{``rather
positive''}, 5 \emph{``very positive'',} 6 \emph{``not applicable /
never had physiotherapy''}{]}. These variables were included as
potential confounders.

To rate the agent's messages, we adapted a scale to measure
\emph{perceived assertiveness} from Babel et al. \cite{babelDevelopmentTestingPsychological2021} (A1-A4) and
created a scale for \emph{perceived expressiveness}, including overall
expressiveness (E1), facial expressiveness (E2) and emotional speech
(E3). To measure \emph{perceived persuasiveness}, we used the
single-item measure \emph{``If I were in this position, I would have
been convinced by the robot''} (P1). We kept the number of items short,
as each scale would be applied 25 times for each participant (see 5.1.3
\emph{Study Design}). The questionnaire utilized a 7-point Likert scale
ranging from \emph{``completely disagree''} to \emph{``completely
agree''} to assess participants' responses. Table 4 presents the rating
scales.

\begin{table}[!htbp]
\caption{Rating Scales for Perceived Assertiveness, Expressiveness, and Persuasiveness\\}
\label{tab:rating-scales}
\renewcommand{\arraystretch}{1.2}
\small
\begin{tabular}{p{0.18\linewidth} p{0.77\linewidth}}
\textbf{Code} & \textbf{Item Text} \\
\midrule

\multicolumn{2}{l}{\textbf{Assertiveness}} \\
A1 & \emph{The message is formulated clearly and directly, without vague or uncertain wording.} \\
A2 & \emph{The message is decisive.} \\
A3 & \emph{The message is motivating.} \\
A4 & \emph{The message demonstrates an appropriate level of assertiveness.} \\
\addlinespace

\multicolumn{2}{l}{\textbf{Expressiveness}} \\
E1 & \emph{The general expressiveness (body language) of the message is adequate.} \\
E2 & \emph{The message uses facial expression adequately.} \\
E3 & \emph{The message uses emotional speech (speed) adequately.} \\
\addlinespace

\multicolumn{2}{l}{\textbf{Persuasiveness}} \\
P1 & \emph{If I were in this position, I would have been convinced by the robot.} \\

\bottomrule
\end{tabular}
\end{table}

\subsubsection{Design}

We employed a five-group within-subject design with five repeated
measures. That is, each participant evaluated all five messages from
each of the five scenarios. This resulted in 25 message ratings per
participant. Our independent variable was agent knowledge with four
factors: \emph{Basic, self-knowledge, user-knowledge,
context-knowledge}. Our dependent variable was \emph{perceived
persuasiveness}. \emph{Perceived expressiveness} and \emph{perceived
assertiveness} were assumed to be mediators.

\subsubsection{Procedure}

Participants gave informed consent and were introduced to the
physiotherapy use case. After filling out a pre-survey, participants
were presented with rating guidelines that explained the relevant
constructs under investigation (\emph{assertiveness} and
\emph{expressiveness}) and the rating procedure. The five scenarios were
then presented in randomized order to account for potential order
effects. For each scenario, all five agent messages had to be rated. The
mean study duration was 58 minutes.

\subsubsection{Analysis}

To assess the robustness of the selected conversations, we conducted a
computational replication analysis using the original model
(ChatGPT-3.5), a newer model from the same platform (ChatGPT-4.1), and a
model from a different platform (Claude 3.5 Haiku). For each model, the
five scenarios were generated 100 times using identical prompts, and
lexical similarity to the original Study 2 conversations was assessed
using TF-IDF cosine similarity. Internal similarity for GPT-3.5 was
moderate to high (0.644--0.730), clearly exceeding a baseline similarity
of 0.296. External similarity was lower but still substantial for
GPT-4.1 (0.558--0.615) and Claude 3.5 Haiku (0.581--0.693). Overall,
these results indicate that the prompts elicit broadly comparable
lexical content across repeated generations and across different LLMs,
supporting the stability and replicability of the scenarios despite
model and output variability.

The reliability and factor structure of the scales to measure
\emph{perceived assertiveness} and \emph{perceived expressiveness} were
assessed using Confirmatory Factor Analysis (CFA) and Omega reliability
coefficients. A multilevel CFA was conducted with observations nested
within participants. The model demonstrated good fit to the data
(\emph{CFI} = 0.962, \emph{TLI} = 0.939, \emph{RMSEA} = 0.073,
\emph{SRMR-within} = 0.058, \emph{SRMR-between} = 0.056), supporting the
proposed two-factor structure. Factor loadings were strong for both
scales, with standardized loadings ranging from 0.78 to 0.89 for
Perceived Assertiveness and from 0.48 to 0.99 for Perceived
Expressiveness. Reliability was evaluated using McDonald's Omega, which
is more suitable for scales with smaller number of items compared to
Cronbach's Alpha \cite{dunnAlphaOmegaPractical2014}, indicating high internal consistency for both
scales (Perceived Assertiveness: $\omega = 0.94$; Perceived
Expressiveness: $\omega = 0.87$).

To test Hypotheses H1-H3, data was transformed into long-format with the
grouping variables being the five scenarios and the five parts of each
scenario. Multilevel structural equation modelling (SEM) was conducted
to test the proposed model structure. This approach is suitable for
mediation analysis
\cite{koschate-fischerMediationAnalysisExperimental2022} and offers
advantages to regression based approaches regarding more detailed
analysis of error terms and more statistical power
\cite{breitsohlANOVAIntroductionStructural2019}, less restrictions
\cite{hayesRegressionbasedStatisticalMediation2017,pekInvalidityTestsSimple2016},
and parsimony \cite{jamesTaleTwoMethods2006}. Multilevel SEM is also useful to
analyze within-subject dynamics throughout repeated measures (e.g., changes of perceptions across
different scenarios)
\cite{newsomStructuralModelsBinary2017}.

Participants were treated as clusters to account for individual
differences, with Level 1 describing within-subject effects and Level 2
describing between-subject effects. In Level 1, all proposed main
effects were modelled, following a 1-1-1 multilevel SEM approach. The
covariance between perceived agent assertiveness and perceived agent
expressiveness was estimated to account for any shared variance not
explained by other predictors. In Level 2, participant effects were
accounted for that influence the dependent variables: \emph{perceived
assertiveness}, \emph{perceived expressiveness}, and \emph{perceived
persuasiveness}. Alternative models were tested by exchanging predictors
with dependent variables and by including confounding variables, such as
message length, participant's age, gender, and previous experiences with
physiotherapy. However, neither of these variables improved the model
fit. In addition, to test the reliability of the SEM approach, linear
mixed models were conducted, with parts being nested within scenarios,
yielding similar parameters and significance levels as with the
multilevel SEM approach.

\subsection{Study 2 -- Results}\label{study-2-results}

To analyze the variable structure of the proposed model, a multilevel
SEM model was estimated using maximum likelihood estimation. The model
included 26 parameters and was based on 675 observations (agent
messages) across 27 clusters (participants). The model yielded a good
fit ($CFI = 0.997$, $TLI = 0.941$, $RMSEA = 0.063$,
$SRMRwithin = 0.008$).

\begin{figure}[ht!]
\begin{center}
\includegraphics[width=\textwidth]{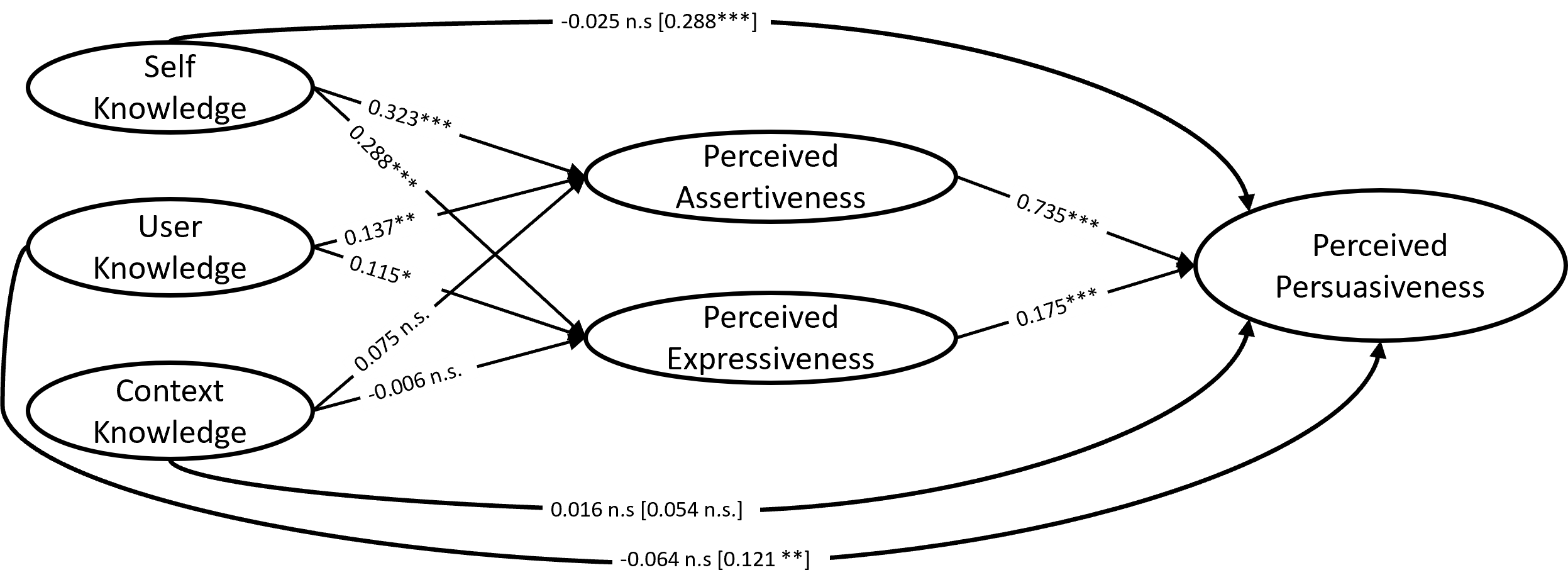}
\end{center}
\caption{Model structure with standardized beta coefficients}
\label{fig:2}
\end{figure}

\emph{Perceived assertiveness} had a large direct effect on
\emph{perceived persuasiveness} ($\beta = 0.735$, $p < .001$).
\emph{Perceived expressiveness} also significantly impacted perceived
persuasiveness, albeit with a small effect size ($\beta = 0.175$,
$p < .001$). \emph{Self-knowledge} did not have a direct effect on perceived
persuasiveness ($\beta = -0.025$, $p = .412$), but showed a medium-sized
indirect effect ($\beta = 0.288$, $p < .001$). Self-knowledge also
impacted \emph{perceived assertiveness} ($\beta = 0.323$, $p < .001$) and
\emph{perceived expressiveness} ($\beta = 0.288$, $p < .001$) with medium
effect sizes. This supports the hypothesized mediation model (H1). \emph{User-knowledge} did not directly impact perceived persuasiveness
($\beta = -0.064$, $p = .516$), but had a small indirect effect
($\beta = 0.121$, $p < .01$). Availability of \emph{user-knowledge} also
led to a small increase in \emph{perceived assertiveness} ($\beta = 0.137$,
$p < .01$) and \emph{perceived expressiveness} ($\beta = 0.115$,
$p < .05$). These findings support the hypothesized mediation (H2). In contrast, \emph{context-knowledge} showed neither a direct effect
($\beta = 0.016$, $p = .570$) nor an indirect effect on \emph{perceived
persuasiveness} ($\beta = 0.054$, $p = .157$). Similarly,
\emph{context-knowledge} had no significant impact on \emph{perceived
assertiveness} ($\beta = 0.075$, $p = .108$) or \emph{perceived
expressiveness} ($\beta = -0.006$, $p = .898$). This contrasts with the
hypothesized mediation (H3).

\subsection{Study 2 -- Discussion}\label{study-2-discussion}

The goal of Study 2 was to investigate whether an agent's available
knowledge led to different perceptions of human raters. Hence, we
conducted an online-study to test whether agent \emph{self-, context-}
and \emph{user-knowledge} impact \emph{perceived assertiveness},
\emph{expressiveness}, and \emph{persuasiveness}. The results speak to
the strong impact of \emph{perceived assertiveness} on \emph{perceived
persuasiveness}, with small effects observed for \emph{perceived
expressiveness}. This aligns with previous research that observed
positive effects of agent assertiveness
\cite{baiLLMgeneratedMessagesCan2025,chidambaramDesigningPersuasiveRobots2012,paradedaWhatMakesGood2019,thomasYouDoorwayNegotiation2018}
and expressiveness
\cite{chenVirtualAugmentedMixed2019,maEnhancingLLMsExpressive2025} on
persuasion. The lower effect sizes in expressiveness in our findings
might stem from the hypothetical nature of the scenarios, as a
physically embodied robot platform could potentially exert a stronger
effect on participants\textquotesingle{} perceptions by expressing
themselves more vividly through vocalization and non-verbal means
\cite{admoniRobotNonverbalBehavior2016,bannaWordsIntegratingPersonality2025,chidambaramDesigningPersuasiveRobots2012,hamCombiningRoboticPersuasive2015}.

The agent's \emph{self-knowledge} -- operationalized through assertive
and expressive characteristics -- had medium sized effects on the
agent's perceived assertiveness and expressiveness. As there was an
indirect, but no direct effect on perceived persuasiveness, perceived
assertiveness and expressiveness fully mediated the effect of
self-knowledge on perceived persuasion effectiveness. This results align
with previous research that found that LLM based GSAs could successfully adapt
their personality
\cite{bannaWordsIntegratingPersonality2025,frischLLMAgentsInteraction2024}.

the agent's \emph{user-knowledge} -- operationalized by the patients'
high age and past profession -- significantly influenced \emph{perceived
assertiveness} and \emph{perceived expressiveness}. Again, a full
mediation was observed, as there was no direct effect on perceived
persuasiveness, but a small indirect effect, supporting H2. However,
these results had less profound effect sizes compared to
\emph{self-knowledge}. The importance of personalization in this study
could have been deemed less relevant as in other research
\cite{lubosLLMgeneratedExplanationsRecommender2024,matzPotentialGenerativeAI2024},
as variations were not strong enough. A reason for that could be that in
our study, raters were unaware of the patient\textquotesingle s actual
age, which represents a methodological limitation. We could have
addressed this issue by providing raters with more comprehensive
profiles of the patient when presenting the scenarios.

\emph{Context-knowledge} -- operationalized by information about
physiotherapy benefits and awareness of time of day -- did not
significantly impact \emph{perceived assertiveness},
\emph{expressiveness,} and \emph{persuasiveness}. This is despite other
research that has observed increased persuasiveness with more topic
knowledge \cite{khanDebatingMorePersuasive2024}. Plausibly, the
manipulation of \emph{context-knowledge} brought some methodological
limitations. First, including time of the day did simply not change the
agent's behavior apart from its greeting, as observed in the qualitative
analysis of Study 1. Second, in conditions that did not explicitly
prompt \emph{context-knowledge}, similar information was already
inherently available to the model -- although potentially less salient.
More specific information that the GSA could not inherently know -- for
example contents of an upcoming physiotherapy session -- could have been
a much more effective manipulation. Future studies could also provide
raters with more contextual information to help them accurately assess
if the agent acted in a responsible manner.

The strong effects of perceived assertiveness and, to a lesser degree,
expressiveness on perceived persuasiveness align with dual-process
models of persuasion such as the Elaboration Likelihood Model (ELM)
\cite{pettyElaborationLikelihoodModel1986}. Although ELM describes
receiver-side processing rather than agent-side message generation, it
predicts that peripheral cues -- such as assertive and expressive
characteristics -- can enhance persuasion when recipients engage in
lower-effort processing. Our findings are consistent with this notion,
as human raters appear to have relied on such cues when judging the
persuasiveness of the agent's messages. Cialdini's
\cite{cialdiniInfluenceSciencePractice2014} principles of authority and
liking provide another potential explanation: assertiveness can function
as a cue for authority, while expressiveness enhances liking and social
presence, both of which are known to strengthen persuasive impact.

\section{General Discussion}\label{general-discussion}

In two exploratory studies, we analyzed persuasive agent messages, and
how the availability of different types of knowledge configurations
might impact a GSA's persuasiveness in a therapy motivation
use case. Results of Study 1 suggest that the availability of
knowledge-prompts may lead to differences in persuasive behavior. Study
2 further supports this, by highlighting that \emph{self-knowledge} and
\emph{user-knowledge} impact user perceptions about the agent's
persuasive behaviors.

Specifically, personality traits included in \emph{self-knowledge}
prompts were found to be particularly important, as corroborated by
existing studies featuring LLM-based agents \cite{frischLLMAgentsInteraction2024}.
Our qualitative analysis further supports this: Assertiveness-related
prompts led to more consistent assertive behavior across scenarios, even
though assertive behavior was also present in scenarios without such
prompts. The qualitative analysis further revealed that the inclusion of
\emph{self-knowledge} about expressive personality traits was essential
for the agent to display emotional expressions effectively.

\emph{User-knowledge} exerted smaller effects on persuasive behaviors,
which may be attributed to the relatively subtle nature of the
manipulations used in this study. For instance, references to past
professions and age might have been insufficiently salient to
significantly influence perceptions. The small effects observed for cues
that relate to the users personal background, suggest that more
personally relevant factors, such as specific preferences or user
personality traits
\cite{gaoAligningLLMAgents2024,lubosLLMgeneratedExplanationsRecommender2024,matzPotentialGenerativeAI2024},
as well as the user's health condition
\cite{vonschallenKnowledgebasedDesignRequirements2026}, could yield
stronger effects. As observed in the qualitative analysis, the agent
utilized user-knowledge effectively when included, especially when
referencing age or past profession. However, these references were not
consistently present across all scenarios, indicating that more detailed
and personalized information might enhance the agent\textquotesingle s
persuasive behavior.

The absence of significant effects for \emph{context-knowledge} aligns
with insights from the qualitative analysis: The time-of-day information
influenced only the greeting behavior, indicating that the manipulation
was not sufficiently impactful. Although the physiotherapy benefits
prompt led to more detailed arguments in some scenarios, the agent
consistently referred to these benefits across most interactions due to
its inherent knowledge of the subject matter. This suggests that more
contextually specific prompts, such as details about the content of a
particular therapy session, may enhance the agent\textquotesingle s
context-sensitive behaviors.

\section{\texorpdfstring{Strengths and Limitations
}{Strengths and Limitations }}\label{strengths-and-limitations}

This paper presents a novel methodological approach to investigate the persausive behavior of GSAs, combining qualitative and quantitative means. As GSAs imitate human behavior based on vast amounts of human-generated
data, their behavioral patterns may be investigated from a psychological
perspective as well. As such, the current research switched the focus
from predefined agent behaviors to observing autonomously generated
persuasive strategies by using fixed human messages as input. While this
method allowed us to compare different knowledge configurations, it also
raises questions on the validity of the interactions for real-life
physiotherapy settings, where human messages have higher variability.
However, the intent behind the scenarios was not to create universal
settings for physiotherapy motivation -- as this would have been
impossible due to near unlimited possibilities in open-ended
interactions. Rather, the goal was to use a specific example with high
relevance as a means to investigate the agent's behavioral patterns in
terms of persuasive strategies and responsible behaviors.

The current research applied a mixed-method approach that not only
analyzed agent behavior qualitatively but also used third-party ratings
of scenarios to evaluate the agent's persuasiveness in an online study.
This had the advantage of high statistical power through within-subject
comparisons, whereas in real-world human-agent interactions, such a
design would have been unrealistic, as participants would have to engage
in the same decision task with multiple variations of the same agent,
which would be tiring and confusing compared to third-party ratings with
more personal distance. However, this approach also introduced a number
of limitations. It is unclear if participants would have actually been
persuaded by the agent, as only perceived persuasiveness was measured.
Further, using text-based scenarios might have impacted experimental
realism for HRI, which could be an explanation that
only small effects for \emph{perceived expressiveness} were
observed\emph{.} In hindsight, creating video-based scenarios in a
similar setting would have made the situation more relatable, which we
highly recommend for future research. The present study provides
prospective means to conduct such research, as well as a basis for
real-world human-agent interaction.

A further limitation lies in the fact that only one specific LLM --
ChatGPT 3-5 -- and one output of this model was used to create the
persuasive scenarios that were used in Study 2. Using the same model
allowed for a better investigation of the effects of different kinds of
available knowledge, but may have hindered generalizability, as other
models -- or an alternative conversation of the same model -- could have
generated different outputs. Hence, we conducted a computational
post-hoc analysis to compare text similarities with different models
(see section 4.1.6). This analysis showed that there are substantial
lexical similarities between different models, but less similarities
than when using the same model. This connects to recent work that
analyzed inner- and inter-LLM writing similarity for thousands of
prompts, finding that texts from the same model are more similar to each
other than to texts from different models
\cite{smithComprehensiveAnalysisLarge2025}. The ChatGPT-3.5 generated
messages used for the scenarios had higher lexical similarities to
messages generated by the same and other LLMs, compared to a baseline
prompt, indicating generalizability of our results. However, future
research should empirically compare different models in terms of their
generated behavior to confirm this.

Despite these limitations we gained valuable insights into behavioral
patterns of persuasive generative agents. The findings not only
contribute to understanding how different knowledge types could
influence the agent's persuasive behavior, but also offer a
methodological foundation that can be applied beyond physiotherapy and
persuasion, potentially guiding future research across diverse domains.
From a broader perspective, insights into knowledge configuration may
support the development of frameworks for responsible generative agents
that respect user autonomy, while effectively promoting desired
behaviors.

\bibliographystyle{plain} 
\bibliography{references}

\end{document}